\documentclass{cpbtex}
\usepackage{graphicx}
\usepackage{siunitx}

\begin{document}

\title{Con-CDVAE: A method for the conditional generation of crystal structures}

\author{
Cai-Yuan Ye$^{1,2}$, \ Hong-Ming Weng$^{1,2,3,}$\thanks{Corresponding author. E-mail:~hmweng@iphy.ac.cn}, and \ Quan-Sheng Wu$^{1,2,}$\thanks{Corresponding author. E-mail:~quansheng.wu@iphy.ac.cn} \\
$^{1}${Beijing National Laboratory for Condensed Matter Physics and Institute of Physics,} \\ {Chinese Academy of Sciences, Beijing 100190, China}\\  
$^{2}${University of Chinese Academy of Sciences, Beijing 100049, China}\\ 
$^{3}${Songshan Lake Materials Laboratory, Dongguan, Guangdong 523808, China}
}   


\date{\today}
\maketitle

\begin{abstract}
In recent years, progress has been made in generating new crystalline materials using generative machine learning models, though gaps remain in efficiently generating crystals based on target properties. This paper proposes the Con-CDVAE model, an extension of the Crystal Diffusion Variational Autoencoder (CDVAE), for conditional crystal generation. We introduce innovative components, design a two-step training method, and develop three unique generation strategies to enhance model performance. The effectiveness of Con-CDVAE is demonstrated through extensive testing under various conditions, including both single and combined property targets. Ablation studies further underscore the critical role of the new components in achieving our model's performance. Additionally, we validate the physical credibility of the generated crystals through Density Functional Theory (DFT) calculations, confirming Con-CDVAE's potential in material science research.
\end{abstract}

\textbf{Keywords:} Crystal structure generation, Machine learning, Conditional generation, Diffusion model  



\section{Introduction}
Material science plays a vital role in the development of modern technology and industry, and materials with excellent properties are the basis for manufacturing a variety of advanced equipment. Crystals, as a kind of material with periodicity, are used in many important fields, such as solar cells, batteries, and catalysis\cite{icsd}.

Although traditional ways, like experiences and first-principles calculations, have catalogued millions of crystals and formed a series of databases, such as the Inorganic Crystal Structure Database (ICSD)\cite{icsd,icsd2}, the Materials Project (MP)\cite{mp}, the Open Quantum Materials Database (OQMD)\cite{oqmd}. But these kinds of methods gradually meet the limitation in searching the complex chemical space and structural space\cite{agrawal2016perspective}.

Fortunately, thanks to the open material databases, some new methods based on machine learning (ML) demonstrated their potential to overcome the limitation. Some ML models have achieved fast prediction of material properties \cite{cgcnn,liang}, reduce computing costs of Density-functional theory (DFT) or molecular dynamics (MD)\cite{MLFF,MLFF2}, which make it possible to combine ML with first-principles calculations and enable efficient search for new materials. The new remarkable work was reported by Deepmind who proposed GNoME model to discover new materials at an astonishing rate\cite{scaling}.

However, combining the discriminative ML models with traditional methods has not completely transcended the limitations of traditional methods. Because they still relied on atomic substitution to explore new materials\cite{scaling}. In the field of natural language and imagery, generative models have shown unexpected creativity while maintaining effectiveness, such as the ChatGPT series\cite{gpt4,gpt3} and the DALL-E series\cite{dalle3,dalle2}. Applying generative modeling to the field of materials searching may break through the limitations of traditional methods fundamentally. At present, the models capable of material generation can be broadly divided into three types. Zekun Ren et al used 2D tables, which were called FTCP representation, to represent crystals and generated new crystals by generating the 2D tables\cite{FTCP}. Kristof T. Schütt et al generated molecular materials by generating atoms one by one\cite{schnet1,schnet2}. Xie T et al combined the graph neural network (GNN), variational autoencoder (VAE), and diffusion model to propose a model called CDVAE for crystal generation\cite{CDVAE}.

Although some progress has been made in crystals generation, there is still a lack of a method that can efficiently generate crystals based on the attributes people need. Table-like representation and generating atoms one by one may not the best way for crystals, so in this article, inspired by DALL-E2, we propose a model based on CDVAE, named Con-CDVAE, which is capable of conditionally generating crystals according to the properties people desire. We design a two-step method to train our model, and generate crystals' latent variables according given properties before generate crystals. We test our model in different conditions (e.g., formation energy, band gap, crystal system, combination of formation energy and band gap), and try different strategies. Finally, we do an ablation experiment for $\textit{Predictor}$ block to demonstrate it can help us to construct the latent variable space.

\section{Method}
\subsection{Diffusion model and crystal generative model}
Diffusion model is a kind of widely used generative model in recent years. It was proposed by Sohl-Dickstein according to the nonequilibrium thermodynamics\cite{diffusion}. Song and Ho respectively improved it and proposed noise conditioned score networks (NCSN\cite{NCSN}) and denoising diffusion probabilistic models (DDPM\cite{DDPM}). The general idea of the diffusion model is to use the diffusion process (also called the forward process) to gradually add noise to the data sample, and gradually transform the data distribution into a prior distribution that is easy to sample, such as Gaussian distribution. In the process, a ML model capable of denoising is trained. In the reverse process, a sample is randomly initialized from the prior distribution, and then the trained model is use to denoise and generate a new sample conforming to the real distribution of the data.

In NCSN\cite{NCSN}, we always set a sequence of increasing standard deviations $\sigma_1 < \sigma_2 < ... < \sigma_T$ and add Gaussian noise to the data.
 \begin{equation}
	\label{eq:noisencsc}
	q(x_t|x_{t-1}) = x_{t-1} + \mathcal{N}(0,(\sigma_t^2-\sigma_{t-1}^2) I),\ q(x_t|x_{0}) = x_{0} + \mathcal{N}(0,\sigma_t^2 I)
\end{equation}
In the reverse process, annealing Langevin dynamics algorithm is used with a ML model called score network $s_\theta(x_t)$.
\begin{equation}
	\label{eq:denoisencsn}
	x_{t-1} = x_t + \frac{\epsilon}{2}\ s_\theta(x_t) + \sqrt{\epsilon}\ \mathcal{N}(0,I)
\end{equation}
where $\epsilon$ is step size and score network $s_\theta(x_t)$ predict the gradient of perturbed data distribution $\triangledown_x \mathrm{log} q(x_t)$.

In DDPM\cite{DDPM}, we always define a sequence of positive
noise scales $0<\beta_1,\beta_2,...,\beta_T$ and get the perturbed data with:
\begin{equation}
	\label{eq:noiseddpm}
	q(x_t|x_{t-1}) = \sqrt{1-\beta_t}\ x_{t-1} + \mathcal{N}(0,\beta_t I),\ q(x_t|x_{0}) = \sqrt{\overline{\alpha}_t}\ x_{0} + \mathcal{N}(0,(1-\overline{\alpha}_t) I)
\end{equation}
where $\overline{\alpha}_t=\prod_{i=1}^t (1-\beta_i)$. And denoise with:
\begin{equation}
	\label{eq:denoiseddpm}
	x_{t-1} = \frac{1}{\sqrt{1-\beta_t}} \left( x_t+\beta_t s_\theta(x_t) \right)  + \sqrt{\beta_t}\ \mathcal{N}(0,I)
\end{equation}

Both diffusion models will be used in this article, where NCSN is used to generate crystals like CDVAE done\cite{CDVAE} and DDPM is used to generate latent variables of crystals. When apply NCSN in crystals generation, we use $M=(A,X,L)$ to represent a crystal. $A, X, L$ represent one-hot vector of atomic type, atomic coordinate and Lattice vector, respectively. In a VAE structure, we use $Encoder(M)$ to get the latent variable $z$ for every crystal, make $Decoder(M_t,z,t)$ act as the scoring network, where $M_t=(X_t,A_t,L)$ means a crystal perturbed by noise at level $t$. We followed the noise addition scheme of CDVAE\cite{CDVAE}:
\begin{equation}
	\label{eq:noiseX}
	X_{t} = X + \mathcal{N}(0,\sigma_t^2 I)
\end{equation}
\begin{equation}
	\label{eq:noiseA}
	A_{t} = \frac{1}{1+\sigma_t} \cdot A + \frac{\sigma_t}{1+\sigma_t} \cdot A_{comp}
\end{equation}
where $A_{comp}$ is the one-hot like vector representing the chemical formula of the crystal. In generation we have
\begin{equation}
	\label{eq:noiseX}
	X_{t-1} = X_t + \eta_t \cdot \epsilon_x + \mathcal{N}(0,\eta_t^2 I)
\end{equation}
where $\eta_t$ is the step size, and $\epsilon_x, A_{t-1}$ is the output of scoring network $Decoder(M_t,z,t)$.

\begin{figure}[h]
    \centering
    \includegraphics[width=0.8\textwidth]{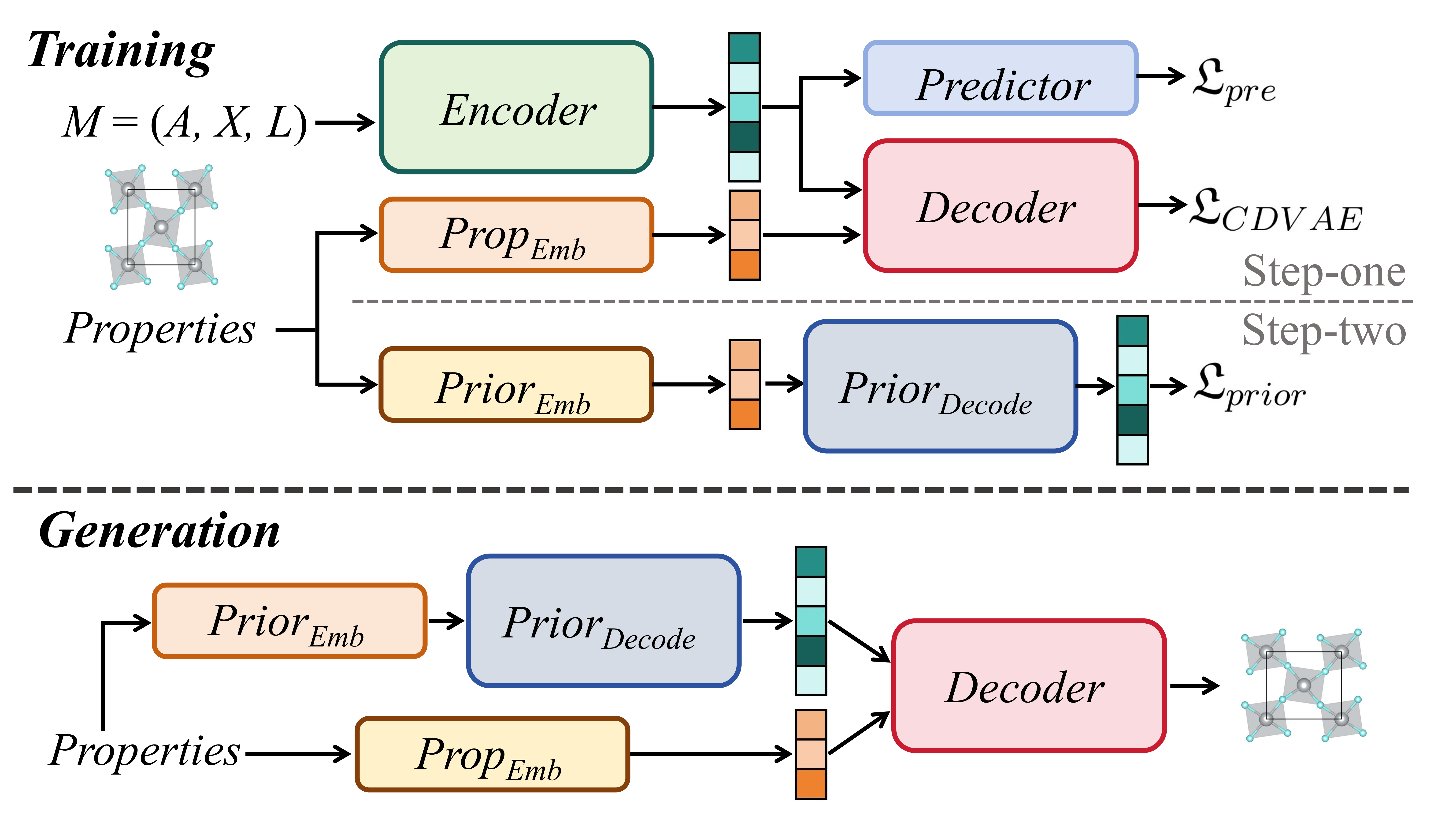}
    \caption{Training and generation flow chart of Con-CDVAE.}
    \label{fig:Con-CDVAE}
\end{figure}

\subsection{Con-CDVAE}

In order to achieve conditional generation, we build two new blocks based on CDVAE\cite{CDVAE}, and introduce the $Prior$ block ,inspired by DALL-E2\cite{dalle2}, to generate the latent variable according given properties. The training and generation flow chart of Con-CDVAE is shown in \autoref{fig:Con-CDVAE} which is worth noting is that we trained the model in two steps. We train some CGCNN models\cite{cgcnn}, a kind of GNN model, to quickly verify whether the generated crystals meet our needs. Although there are errors in using ML models for verifying, it can also meet the needs of preliminary verification.

\subsubsection{Condition.}

As shown in \autoref{fig:Con-CDVAE}, we build $Prop_{Emb}$ and $Predictor$ blocks based on the VAE structure of CDVAE\cite{CDVAE}. $Prop_{Emb}$ block embeds the properties, and there are differences in how continuous and discrete properties are embedded. Continuous properties are expanded by Gaussian basis function and embedded by an MLP. 
\begin{equation}
	\label{eq:con_emb}
	C_{con} = \mathrm{MLP}\left( \left[e^{-\frac{(a_{con}-a_{min})^2}{2\sigma^2}}, e^{-\frac{(a_{con}-a_{min}-\sigma)^2}{2\sigma^2}},...,e^{-\frac{(a_{con}-a_{max})^2}{2\sigma^2}}\right]\right)
\end{equation}
where $a_{con}$ is property value, $a_{min}$ is the minimum, $a_{max}$ is the maximum, and $\sigma$ is grid spacing. Each class of discrete properties are represented by a learnable vector and then processed with an MLP.
\begin{equation}
	\label{eq:dis_emb}
	C_{dis} = \mathrm{MLP}\left( a_{dis}^{k}\right)
\end{equation}
where $a_{dis}^{k}$ is the learnable vector of k-th class of the discrete property. And we apply an MLP to mix embedding vectors of different properties when train the model with combination contditon. Then the latent variable $z$ and properties embedding will be mix by an MLP before fed into the $Decoder$. 
\begin{equation}
	\label{eq:dis_emb}
	C_{prop} = \mathrm{MLP}\left( C_{a_1} \oplus C_{a_2} \oplus ... \oplus C_{a_n}\right)
\end{equation}
\begin{equation}
	\label{eq:z_con}
	z_{con} = \mathrm{MLP}\left( C_{prop} \oplus z\right)
\end{equation}
where $\oplus$ represents concatenation of vectors.

The $Predictor$ block is an improvement of the property predictor in CDVAE\cite{CDVAE}, which can only be applied to one continuous property. $Predictor$ block can be applied to multiple properties at the same time, including discrete properties. This block use latent variable $z$ as input to make crystals with similar properties close in the latent space which may be helpful when use $Prior$ to generate new latent variable with properties. The normalized mean squared error (MSE) loss function is used for continuous properties and the cross entropy loss function is used for discrete properties.
\begin{equation}
	\label{eq:Lpre}
	\mathfrak{L}_{pre} = \left( \widehat{a}_{con}-\widetilde{a}_{con} \right)^2 + \mathrm{CrossEntropy}(\widehat{a}_{dis},a_{dis}), \widetilde{a}_{con}=\frac{a_{con}-a_{min}}{a_{max}-a_{min}}
\end{equation}
where $\widehat{a}_{con}$ and $\widehat{a}_{dis}$ represent predicted values. Using the joint loss function $\mathfrak{L}_{joint}=\mathfrak{L}_{CDVAE}+\alpha \cdot \mathfrak{L}_{pre}$, we train $Encoder, Predictor, Prop_{Emb}, Decoder$ first, which is call "Step-one" training. The hyperparameter $\alpha$ is usually set to 3.

\subsubsection{Prior.}

After the "Step-one", we build the $Prior$ block, which is the DDPM version of the diffusion model. The task of $Prior$ is to generate latent variables with the given properties, so we use the result of $Encoder$ as label. $Prior$ can be divided into two sub-blocks: $Prior_{Emb}$ and $Prior_{Decode}$. $Prior_{Emb}$ is similar with $Prop_{Emb}$, but can embed chemical formula when we need.
\begin{equation}
	\label{eq:chemical}
	C_{che} = \mathrm{MLP}\left( n_{H}\cdot v_{H} \odot  n_{He}\cdot v_{He} \odot ...\right)
\end{equation}
where $n_{H}$ is the fraction of hydrogen, $v_{H}$ is the 92-dimensional embedding vector of hydrogen, and $\odot$ means adding the elements of the corresponding positions of two vectors. The 92-dimensional embedding vector of elements is got from CGCNN\cite{cgcnn}. $Prior_{Decode}$ consists of six layers of neurons. The number of neurons in the first three layers decreases layer by layer, and the number of neurons in the last three layers increases layer by layer. There is a residual structure between the layers with the same number of neurons. We use Gaussian noise $z^t= \sqrt{\overline{\alpha}_t} z+ \sqrt{1-\overline{\alpha}_t} \epsilon_z$, where $\epsilon_z \sim \mathcal{N}(0, I)$. We use the loss function $\mathfrak{L}_{prior}$ to train $Prior$, where $\hat{\epsilon}_z=Prior_{Decode}(z^t, C_{prop}, t)$.

\begin{equation}
	\label{eq:Lprior}
	\mathfrak{L}_{prior} = \mathbb{E}_{\epsilon_z \sim \mathcal{N}(0, I), t\sim \mathcal{U}(0,T)}\left \| \hat{\epsilon}_z-\epsilon_z \right \|^2 
\end{equation}

For the training of the $Prior$ block ("Step-two" training process), we divided it into two approaches. The first one involves training with properties identical to those of the $Prop_{Emb}$ block's input, referred to as the default condition $Prior$. The second one involves training with properties such as band gap, formation energy, crystal system, space group, convex hull energy, atomic number, and chemical formula, termed as the full condition $Prior$. For example, if we want to train a Con-CDVAE capable of crystal generation based on band gap as a condition, then the input for the $Prop_{Emb}$ block will only include band gap. The input for the default condition $Prior$ will also only include band gap, while the input for the full condition $Prior$ will include all the mentioned properties.

Using these two kinds of $Prior$, we proposed three strategies during the generation process. The first one involves generating latent variables using the default condition $Prior$, referred to as the $default$ strategy. The second one involves utilizing the full condition $Prior$ and providing all the required properties as input, referred to as the $full$ strategy. The third one involves using the full condition $Prior$ but inputting only a subset of properties referred to as the $less$ strategy. The missing attributes are randomly sampled from the training set based on the provided properties to complete the model's input. This three strategies will be labeled as "D", "F", "L" in the following figures and tabels.

After generating latent variables using the $Prior$, we employed the $Predictor$ block to filter the generated latent variables, selecting those that best meet our predefined criteria to input into the $Decoder$ for crystal generation. The filtering ratio is 20000:200, meaning that for each test mentioned in \autoref{sec:experiments}, two hundred crystals were generated. Similar methods can also be observed in DALL·E2\cite{dalle2}.

\begin{table}[]
	\centering
	\renewcommand{\arraystretch}{1.0}
	\caption{The generation success rate of models trained with different data sets using different strategies is obtained by using different classification of band gap and formation energy as generation conditions}
	\begin{tabular}{ccccc}
		\hline 
		  & Regression BG &Regression FE &Classification of BG &Classification of FE \\
            & MAE(eV)$\downarrow$  &MAE(eV/atom)$\downarrow$ &Accuracy(\%)$\uparrow$ &Accuracy(\%)$\uparrow$ \\
		\hline 
		MP20  &   0.362 & 0.095 & 86.9 &97.0 \\ 
		MP40  &   0.359 & 0.081 &86.9 & 98.4  \\ 
		OQMD  &   0.124 & 0.059 & 95.6 &98.5\\  
		\hline 
	\end{tabular}
	\label{tab:cgcnn}
\end{table}

\subsection{Evaluation of the generated crystals}

We use \texttt{pymatgen} Python package\cite{pymatgen} and several trained CGCNN models to quickly validate the generation performance of the model. We analyze the crystal system of the generated crystals via \texttt{SpacegroupAnalyzer} utility from the \texttt{pymatgen} with the parameters: \texttt{symprec=0.2, angle\_tolerance=5}. Two regression models were trained separately utilizing band gap and formation energy. Two classification models were trained with thresholds based on whether the band gap is 0 eV and whether the formation energy is greater than -0.9 eV/atom. These two thresholds are selected based on the data distribution of the MP database. The performance of CGCNN on the test set is shown in \autoref{tab:cgcnn}. Although there are errors in the verification based on machine learning model, it is sufficient to reflect the ability of Con-CDVAE to generate crystals according to attributes, and it can be preliminatively verified before further doing a large number of DFT calculations.

\section{Experiments}
\label{sec:experiments}

\label{subsec:single}
\begin{figure}[h]
	\centering
	\includegraphics[width=1.\textwidth]{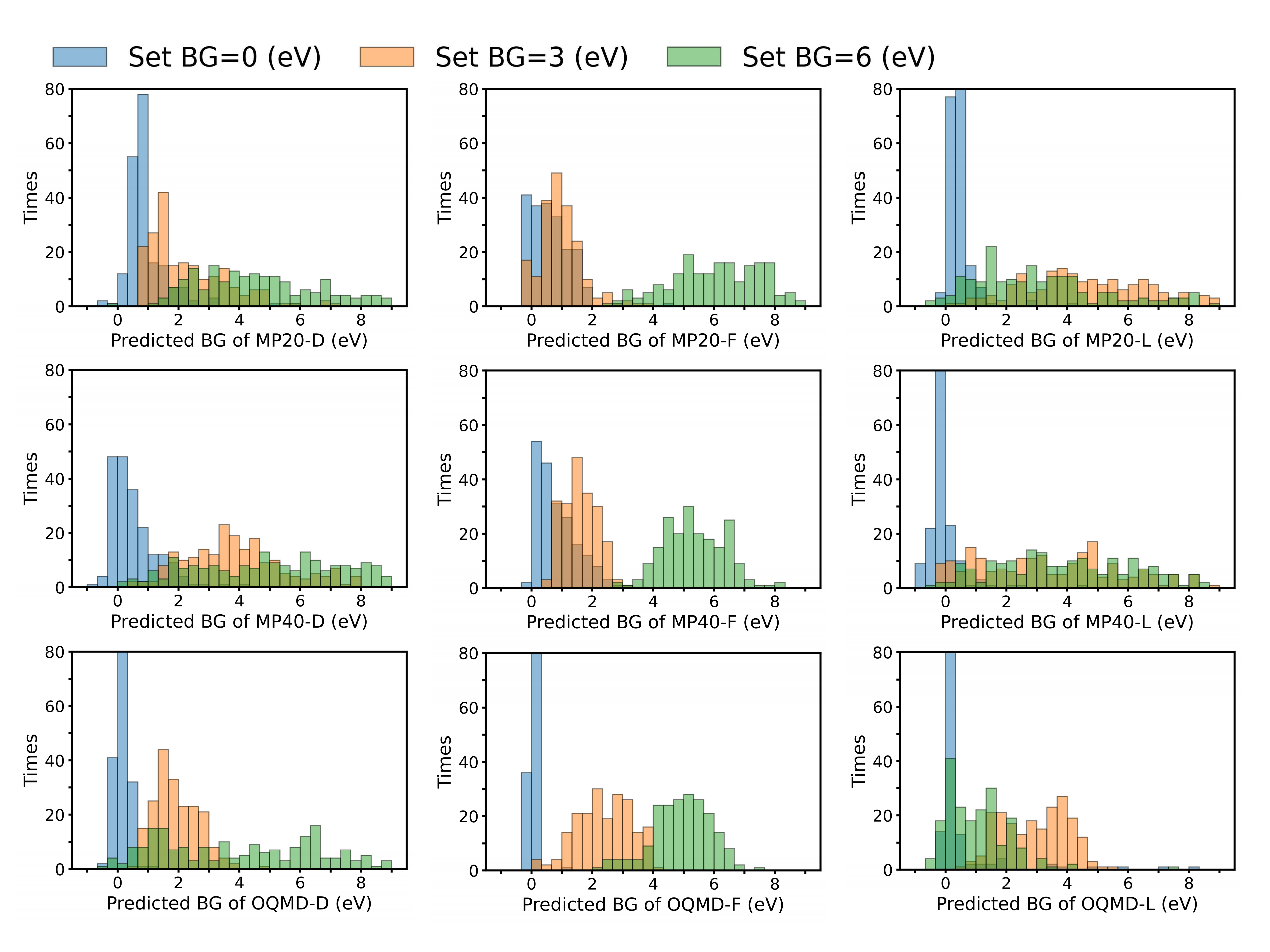}
	\caption{Predicted band gap distributions of crystal generated by Con-CDVAE. The first to third lines are the results obtained by training with the MP20, MP40, and OQMD datasets, respectively. The first to third columns are crystals generated using the $default$, $full$, and $less$ strategy respectively.}
	\label{fig:bandgap}
\end{figure}

To train Con-CDVAE, we use the data get from MP and OQMD, and filter into 3 sub-datasets. We first filtered by atomic number density greater than 0.01 atom/$\si{\angstrom}^3$, energy above hull less than 0.5 eV/atom and band gap less then 8 eV. Subsequently, selecting cells with atomic numbers not exceeding 20, 40 and 20 we obtain 3 sub-datasets called MP20, MP40, and OQMD20 (For simplicity, we will refer to it as OQMD in the following text.). There are 71665 crystals in MP20, 108039 crystals in MP40, and 616412 crystals in OQMD. Distribution of band gap, formation energy and crystal system in three sub-datasets are shown in \autoref{fig:distribution}. We observed that, in the three sub-datasets, materials with zero bandgap constitute a significant proportion. The formation energy distribution of MP20 and MP40 exhibits two peaks around 0 eV/atom and -2.5 eV/atom, while OQMD shows a unimodal distribution around 0 eV/atom. The crystal system distribution of MP20 and MP40 is more uniform compared to OQMD. Cubic crystal structures are more prevalent in OQMD. We utilized these three subdatasets to train Con-CDVAE with a ratio of 8:1:1 for training set, validation set, and test set, respectively.

\subsection{Generating with single condition}

\begin{figure}[h]
	\centering
	\includegraphics[width=1.\textwidth]{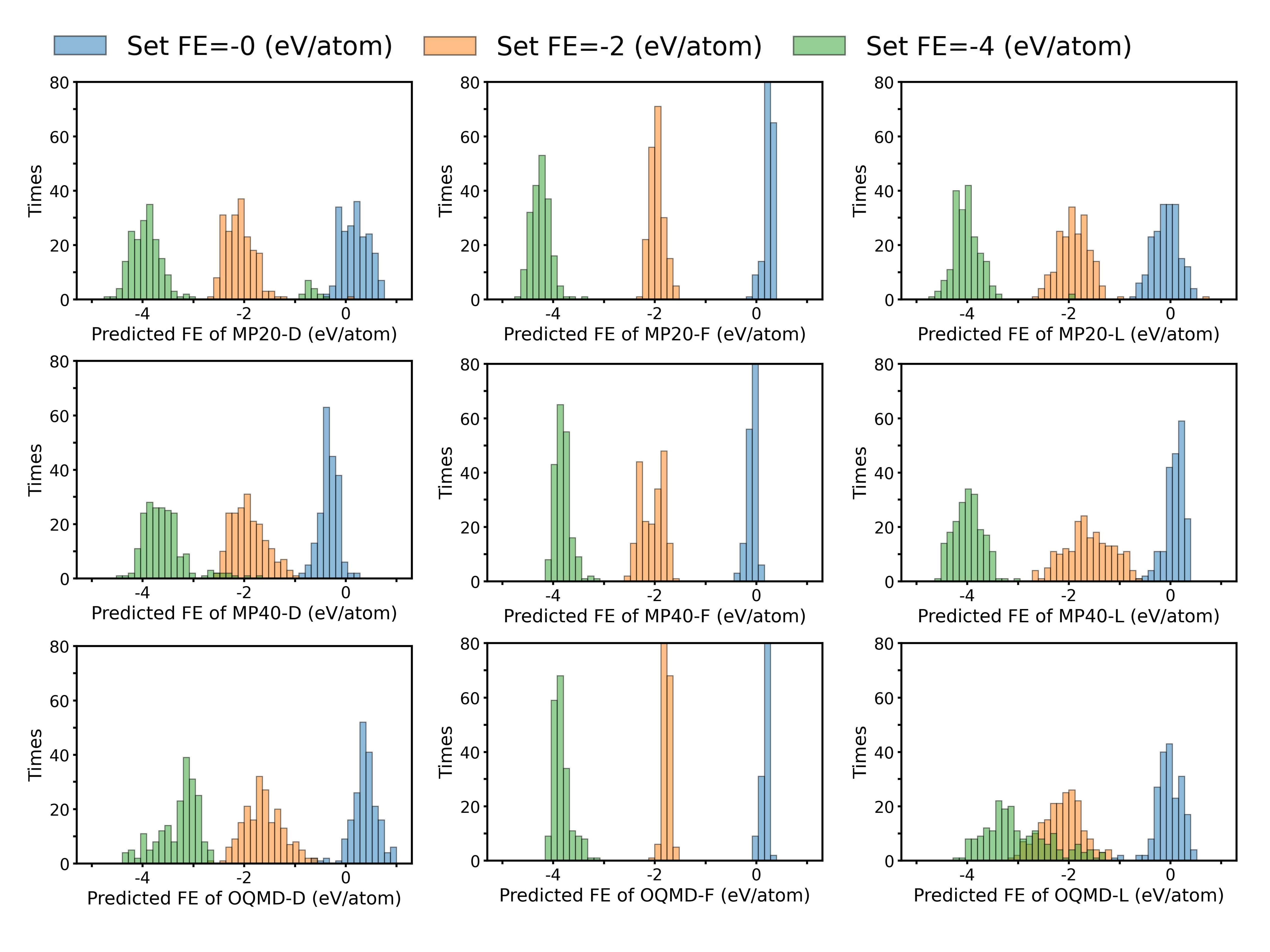}
	\caption{Predicted formation energy distributions of crystal generated by Con-CDVAE. The first to third lines are the results obtained by training with the MP20, MP40, and OQMD datasets, respectively. The first to third columns are crystals generated using the $default$, $full$, and $less$ strategy respectively.}
	\label{fig:format}
\end{figure}

We initially trained Con-CDVAE separately using band gap (BG), formation energy (FE), and crystal system (CS) as single condition. Three independent models were trained on the three sub-datasets, and all were tested using three strategies. When using the $less$ strategy for crystal generation based on crystal system as condition, we only input the crystal system into the full condition $Prior$. However, when generating based on band gap or formation energy as condition, we only input band gap and formation energy as conditions, with the missing conditions randomly sampled from the training set.  \autoref{fig:bandgap} illustrates our crystal generation based on band gap of 0 eV, 3 eV, and 6 eV (represented by blue, yellow, and green bars, respectively). The bandgap distributions are then depicted in histogram form. Different subplots represent different datasets and strategies, where the first to third rows correspond to MP20, MP40, OQMD, and the first to third columns correspond to the $default$ strategy, $full$ strategy, and $less$ strategy. \autoref{fig:format} shows our crystal generation based on formation energy of 0 eV/atom, -2 eV/atom, and -4 eV/atom in the similar format.

\begin{figure}[h]
	\centering
	\includegraphics[width=1.\textwidth]{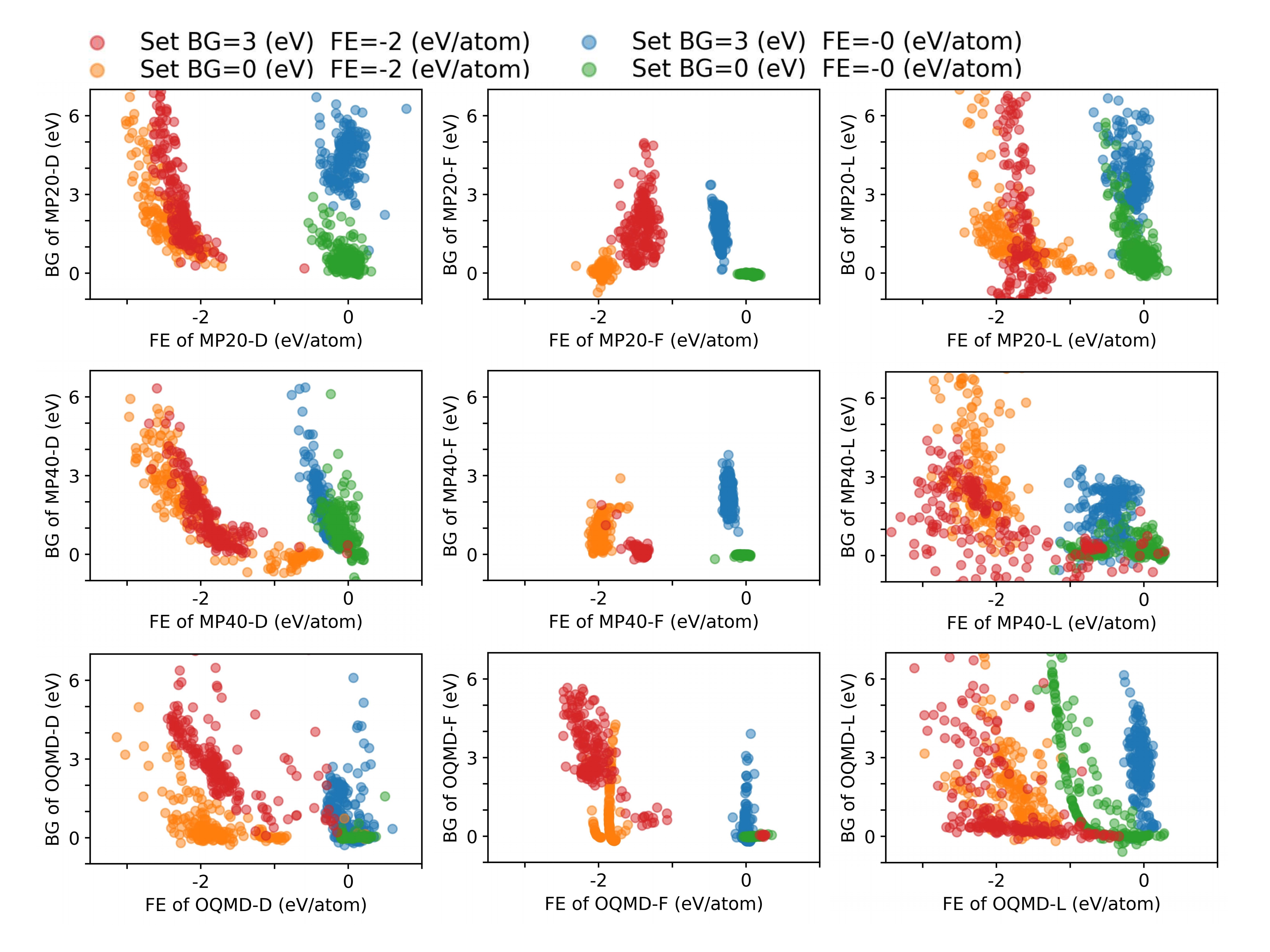}
	\caption{Predicted band gap and predicted formation energy of crystals generated by Con-CDVAE. The first to third lines are the results obtained by training with the MP20, MP40, and OQMD datasets, respectively. The first to third columns are crystals generated using the $default$, $full$, and $less$ strategy respectively.}
	\label{fig:FEBG}
\end{figure}

All three single condition tests indicate that the results generated using the $full$ strategy are most in line with our set condition, achieving the intended purpose of designing the full strategy. However, the $full$ strategy limits the application of our model. In practical applications, we often only have requirements for one or a few properties, and cannot completely obtain all the information of an unknown crystal. Therefore, we design the $less$ strategy. However, from the test results, the $less$ strategy based on random extraction cannot maintain the same capability as the $full$ strategy, and is not significantly better than the $default$ strategy. We believe that increasing the amount of training data can enhance the model's performance, as evident in the $default$ strategy for crystal system and the $full$ strategy for formation energy. However, the distribution and quality of the data itself can also impact model training. For instance, in the OQMD dataset, a higher proportion of cubic crystal system materials leads to better performance when generating crystals with a cubic crystal system, compared to models trained on the other two databases. Additionally, OQMD has a lower proportion of materials with formation energy less than -2 eV/atom, resulting in slightly poorer performance when generating crystals with a formation energy of -4 eV/atom compared to models trained on the other two databases.

Comparing the results from the three types of tests, we find that the generation results based on the condition of formation energy are superior to those based on band gap, which, in turn, are better than those based on crystal system. This is because the mapping from chemical and structural information to formation energy is simpler than the mapping to band gap, and the corresponding inverse mapping is also simpler. The inferior results in crystal system generation are attributed to our use of the \texttt{pymatgen} library for crystal system determination, which imposes stricter tolerance conditions. Crystal structures generated directly using the diffusion model without undergoing DFT relaxation may introduce errors that exceed this tolerance threshold.

Although Con-CDVAE generates better results in regions with sufficient training data, we can still see that it has the ability to perform conditional generation in regions with less data, such as the result generated with a band gap of 3 eV.

We also try to generate crystals with crystal system as condition. The result is show in Appendix B \autoref{tab:crystal-system}, which show that Con-CDVAE is not very good at generating the structure of fixed space groups, because CDVAE cannot guarantee the symmetry of the generated crystals well. Further relaxation with DFT may be needed to better show the symmetry of the generated crystals, which is also an area that needs to be perfected in the future.

\subsection{Generating with combination condition}

Then, we use band gap and formation energy as combination conditions to train Con-CDVAE to test the generation ability of the model under multiple conditions. The results are shown in \autoref{fig:FEBG}, where each point represents a generated crystal, and the horizontal and vertical coordinates of the points represent the formation energy and band gap predicted by CGCNN respectively. Points of different colors represent different conditions set during generation. Red indicates that the band gap is set to 3 eV and the formation energy is set to -2 eV/atom. Yellow indicates that the band gap is set to 0 eV and the formation energy is set to -2 eV/atom. Blue indicates that the band gap is set to 3 eV and the formation energy is set to 0 eV/atom. Green indicates that the band gap is set to 0 eV and the formation energy is set to 0 eV/atom.

As can be seen from the figure, Con-CDVAE has a stronger ability to generate crystals with a specific formation energy than crystals with a specific band gap (the data points are distributed longitudinally). By comparing \autoref{fig:bandgap} and \autoref{fig:FEBG}, we find that combination condition generation will degrade the performance of model, but the model still has certain ability to generate with combination condition, especially when the $full$ strategy is used.

In order to reduce the difficulty of generation, we designed a joint conditional generation test with discrete attributes using formation energy and bandgap. The specific method involved labeling crystals with zero bandgap as 'zero' and the rest as 'non-zero', while crystals with formation energy greater than -0.9 eV/atom were labeled as 'high' and the rest as 'low'. The model was trained based on these labels, and the final results are shown in \autoref{tab:foramt-bandgap}.

\begin{figure}[h]
	\centering
	\includegraphics[width=1.\textwidth]{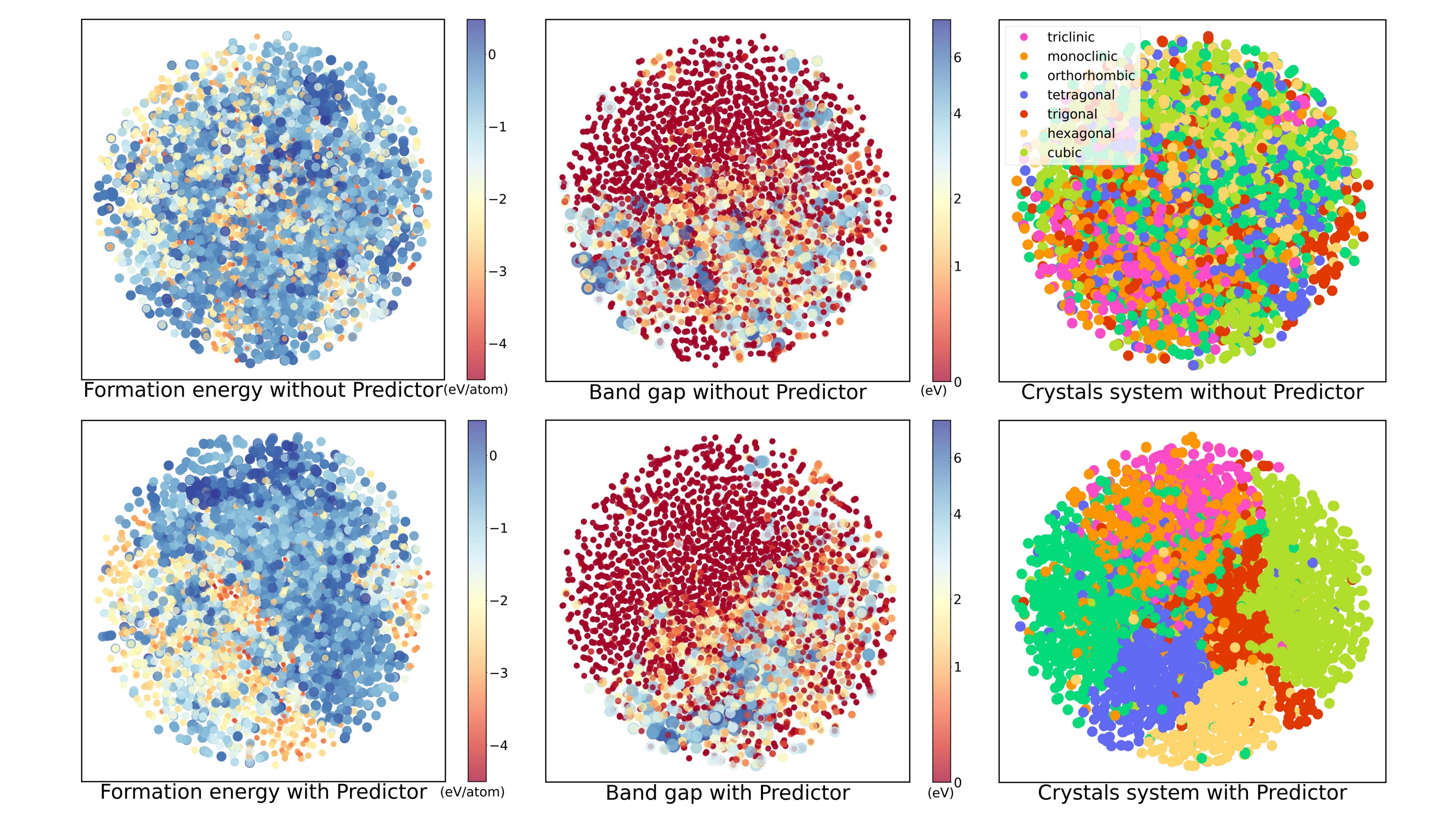}
	\caption{T-SNE of crystals' latent variables which are obtained by the Con-CDVAE. The first line obtained by training with the $Predictor$ block, the second line without the $Predictor$ block. The first to third columns are the results obtained by training with formation energy, band gap, and crystal system as condition, respectively.}
	\label{fig:tsne}
\end{figure}

\begin{table}[]
	\centering
	\renewcommand{\arraystretch}{1.0}
	\caption{The generation success rate of models trained with different data sets using different strategies is obtained by using different classification of band gap and formation energy as generation conditions}
	\begin{tabular}{ccccc}
		\hline 
		target  & BG:zero; &BG:zero; &BG:non-zero; &BG:non-zero; \\ 
		classification  & FE:high (\%) &FE:low (\%) & FE:high (\%) &FE:low (\%) \\
		\hline 
		MP20-D  &   0.0 & 0.0 & 53.5 &99.5 \\ 
		MP20-F  &   100.0 & 74.0 &99.5 & 22.0  \\ 
		MP20-L  &   92.5 & 5.0 & 96.0 &99.5\\ 
		\hline 
		MP40-D  &   90.5 & 11.0 & 90.0 &99.0 \\ 
		MP40-F  &   100.0 & 51.5 & 43.5 &100.0 \\ 
		MP40-L  &   90.0 & 60.0 & 99.0 &98.5\\  
		\hline 
		OQMD-D  &   88.5 & 37.0 & 27.0 &92.0 \\ 
		OQMD-F  &   96.5 &90.5 & 46.0 &39.0 \\ 
		OQMD-L  &   88.5 & 16.0 & 60.0 &62.0\\  
		\hline 
	\end{tabular}
	\label{tab:foramt-bandgap}
\end{table}

\subsection{Ablation experiment for $\textit{Predictor}$ block}

Using MP20, we train three Con-CDVAE without the $Predictor$ block on formation energy, band gap, and crystal system as the single condition, which means setting the hyperparameter $\alpha$ in the joint loss function $\mathfrak{L}_{joint}$ to 0. After the training, $Encoder$ is used to obtain the latent variables of the crystals in test set, and T-sne\cite{tsne} is used for dimensionality reduction and two-dimensional dot plot is drawn. Then, corresponding two-dimensional point plots are drawn using the three MP20-trained models mentioned in \autoref{subsec:single}, and the results are shown in \autoref{fig:tsne}. 

The three subplots in the first row of the figure depict the results without the $Predictor$ block, while the second row shows the results with the $Predictor$ block. In the plots of the first column, the closer the color is to blue, and the larger the size of the points, the higher the formation energy of the corresponding crystals. In the plots of the second column, the closer the color is to blue and the larger the size of the points, the larger the band gap of the corresponding crystals. Here, the logarithm is taken for the band gap as most crystals have a band gap of 0. In the plots of the third column, points of different colors represent crystals of different crystal systems. As we can see, $Predictor$ effectively aggregates crystals with similar properties in the latent space, which is important for structuring latent space.

\subsection{Validation by DFT}

\begin{figure}[h]
	\centering
	\includegraphics[width=0.5\textwidth]{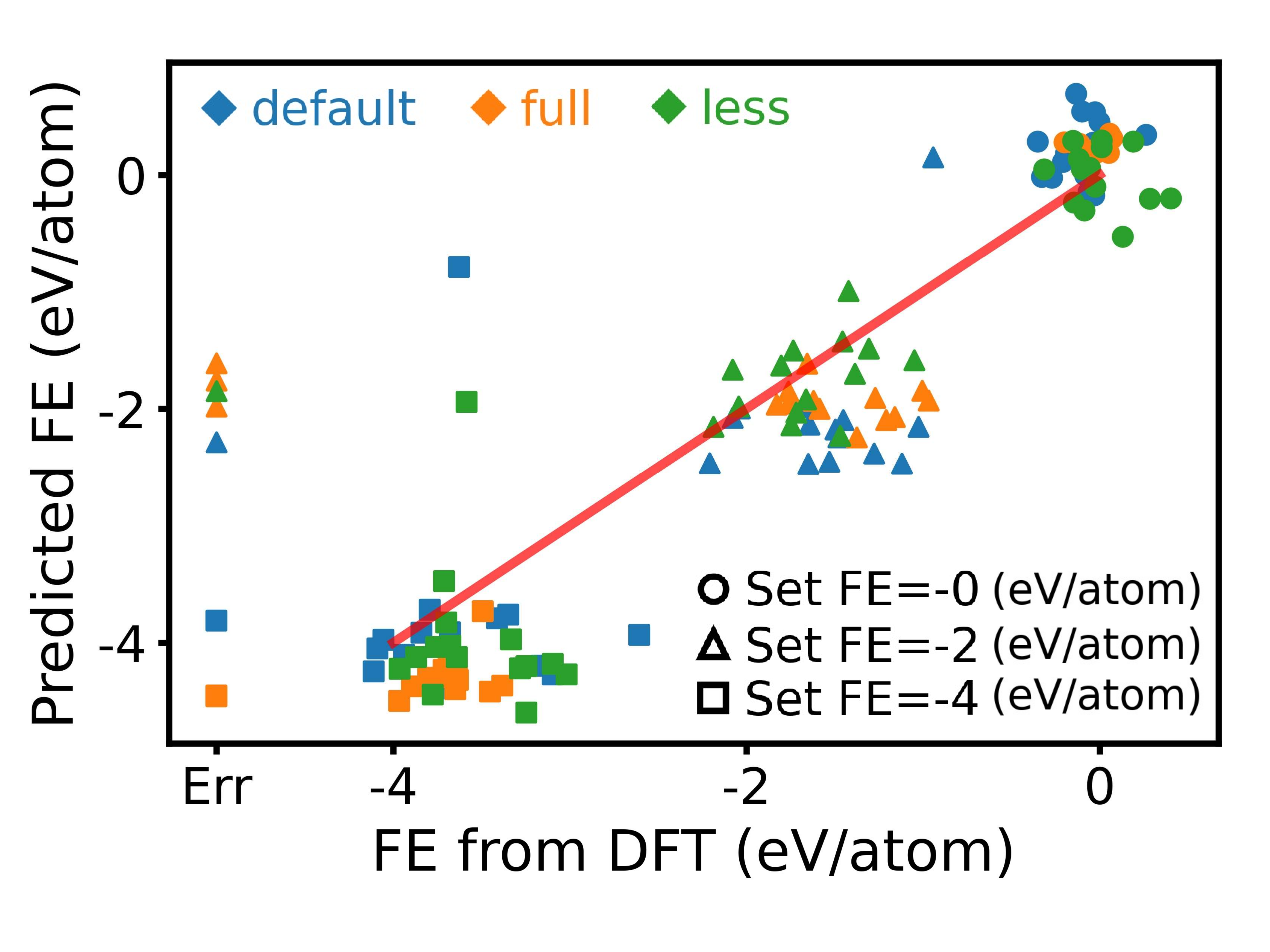}
	\caption{The predicted formation energy and the 
calculated formation energy of 135 crystals generated by Con-CDVAE. The blue, yellow, and green dots represent crystals generated using the $default$, $full$, and $less$ strategy, respectively. The points of the circle, triangle, and square represent the target formation energy 0, -2, and -4 eV/atom, respectively, set during generation.}
	\label{fig:relax}
\end{figure}

In order to verify the rationality of crystals generated by Con-CDVAE, We randomly selected 135 model-generated materials for relaxation by DFT in the test with formation energy as single condition (Three formation energy conditions and three strategies constitute nine cases, and 15 crystals are randomly selected in each case). At the same time, we also calculate formation energy of them.

The result is shown in \autoref{fig:relax}. The vertical coordinate is the formation energy predicted by CGCNN, and the horizontal coordinate is the formation energy calculated by DFT, and the eight crystals that fail to relax are labeled "Err". Dots of different shapes represent crystals generated with different formation energy conditions, and dots of different colors represent different generation strategies. The DFT calculation further verifies the ability of Con-CDVAE to generate crystals according to the conditions.

As shown in \autoref{tab:abc}, we calculate the mean absolute change (MAC) of lattice constants and lattice angles between initial generated structures and DFT relaxed structures and use \texttt{StructureMatcher} from \texttt{pymatgen} to calculate root mean squared displacement (RMSD).

\begin{table}[]
	\centering
	\renewcommand{\arraystretch}{1.0}
	\caption{The MAC of lattice constants and lattice angles and RMSD crystals before and after relaxation.}
	\begin{tabular}{ccccccc}
		\hline 
		MAC$_a(\si{\angstrom})$& MAC$_b(\si{\angstrom})$ &MAC$_c(\si{\angstrom})$ &MAC$_{\alpha}(\si{\angstrom})$ &MAC$_{\beta}(\si{\angstrom})$&MAC$_{\gamma}(\si{\angstrom})$&RMSD$(\si{\angstrom})$ \\
		  0.376&0.397&0.740&4.694&4.727&3.818&0.307\\
		\hline 
	\end{tabular}
	\label{tab:abc}
\end{table}

\begin{figure}[h]
	\centering
	\includegraphics[width=1.\textwidth]{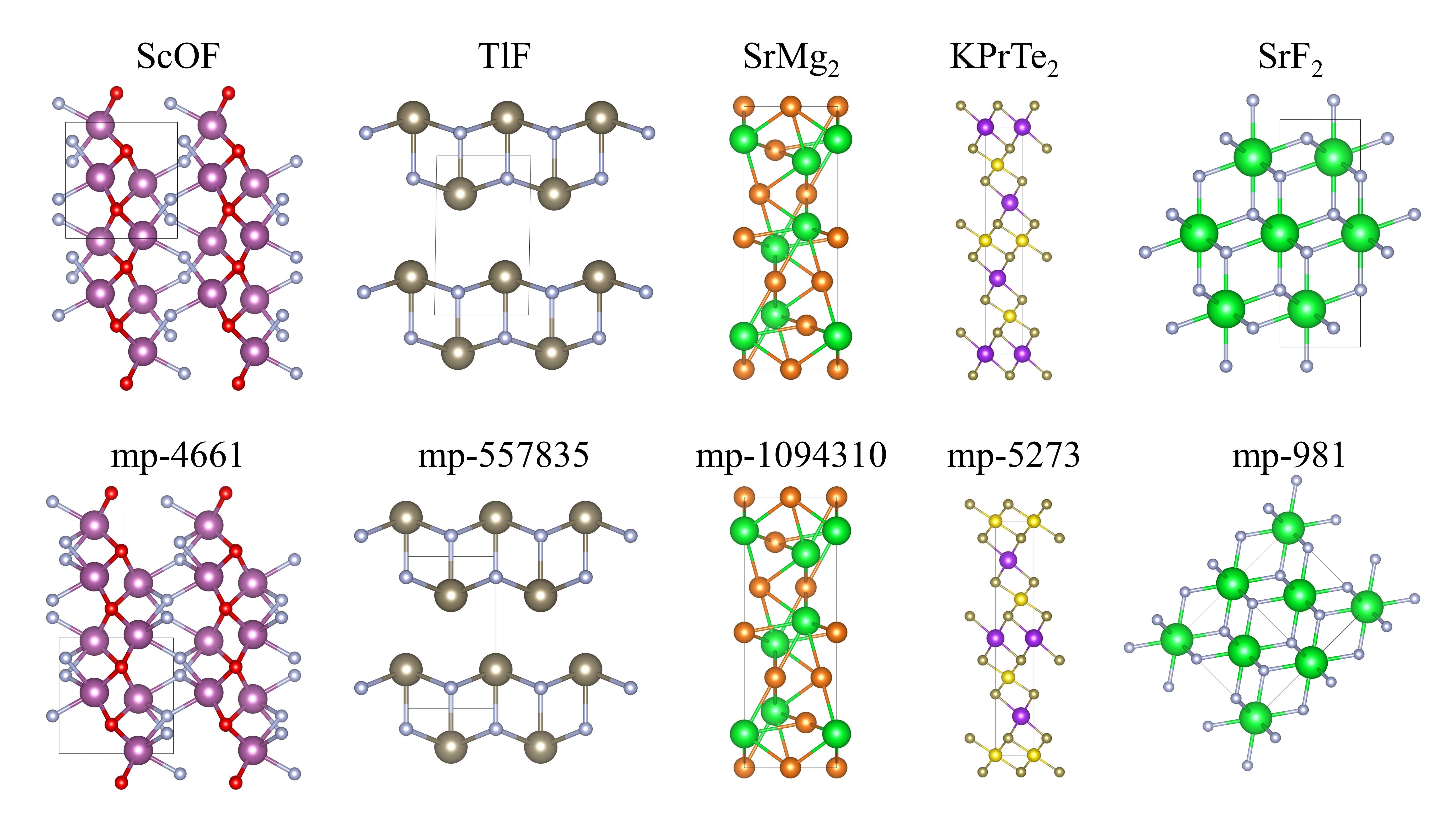}
	\caption{Five structure generated by Con-CDVAE are similar with crystals from MP\cite{mp}. The second row of crystals is derived from MP, and the corresponding ID is their ID in the MP database.}
	\label{fig:gen}
\end{figure}

\begin{figure}[h]
	\centering
	\includegraphics[width=1.\textwidth]{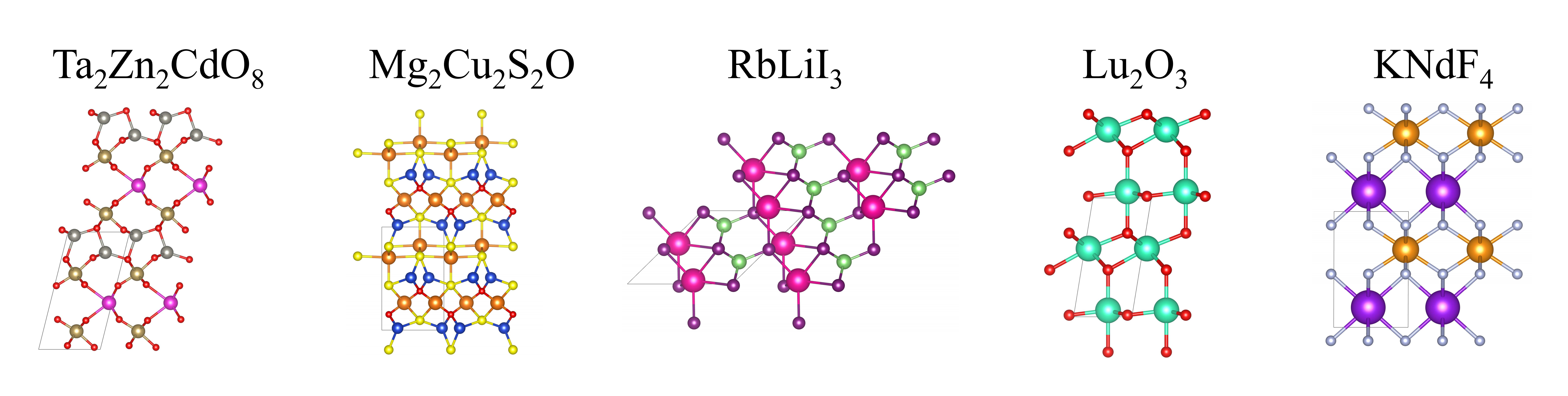}
	\caption{Five structures generated by Con-CDVAE did not find any similar crystals in the MP database.}
	\label{fig:gen2}
\end{figure}

For the 127 crystals that successfully relaxed, we use \texttt{StructureMatcher} with the following default parameters: \texttt{ltol=0.2, stol=0.3, angle\_tol=5} to find out if similar materials already existed in MP. We found a total of 22 pairs, and \autoref{fig:gen} shows 5 of them. Notably, mp-981 does not appear in the training set. This shows that our model has the ability to generate crystals outside of the training set. In \autoref{fig:gen2} we show five gennerated crystals which do not find matching materials in MP.

\section{Conclusion}

In this paper, a diffusion model, Con-CDVAE, which can effectively realize crystal generation according to target properties, is proposed. We design a two-step model training method and three generation strategies. We test the model under a variety of conditions, including single and combination conditions, and checked the results with ML model. We found that although the model performs well in regions with sufficient data, it can also generate crystals in regions with insufficient data. Then, we perform preliminary ablation experiments on the model and demonstrate the importance of the $Predictor$ block for building the spatial structure of latent variables. Finally, we performed a simple DFT validation, further confirming the rationality of the crystals generated by Con-CDVAE, as well as the model's ability to generate crystals based on target properties and to generate crystals outside the training set.

Of course, there are areas in our work that can be further improved. The model needs to be further improved to recognize the symmetry information of crystals and realize the generation of crystals with specific space groups. How to improve the model to reduce the MAC and RMSD of the generated crystals before and after relaxation requires further investigation. Other more practical applications are physical properties that are also worth trying further, such as the critical temperature of superconducting materials.

\newpage
\appendix

\addcontentsline{toc}{chapter}{Appendix A: Data distribution}
\renewcommand\thefigure{A\arabic{figure}} 
\section*{Appendix A: Data distribution}
\setcounter{figure}{0}
\begin{figure}[h]
	\centering
	\includegraphics[width=1.\textwidth]{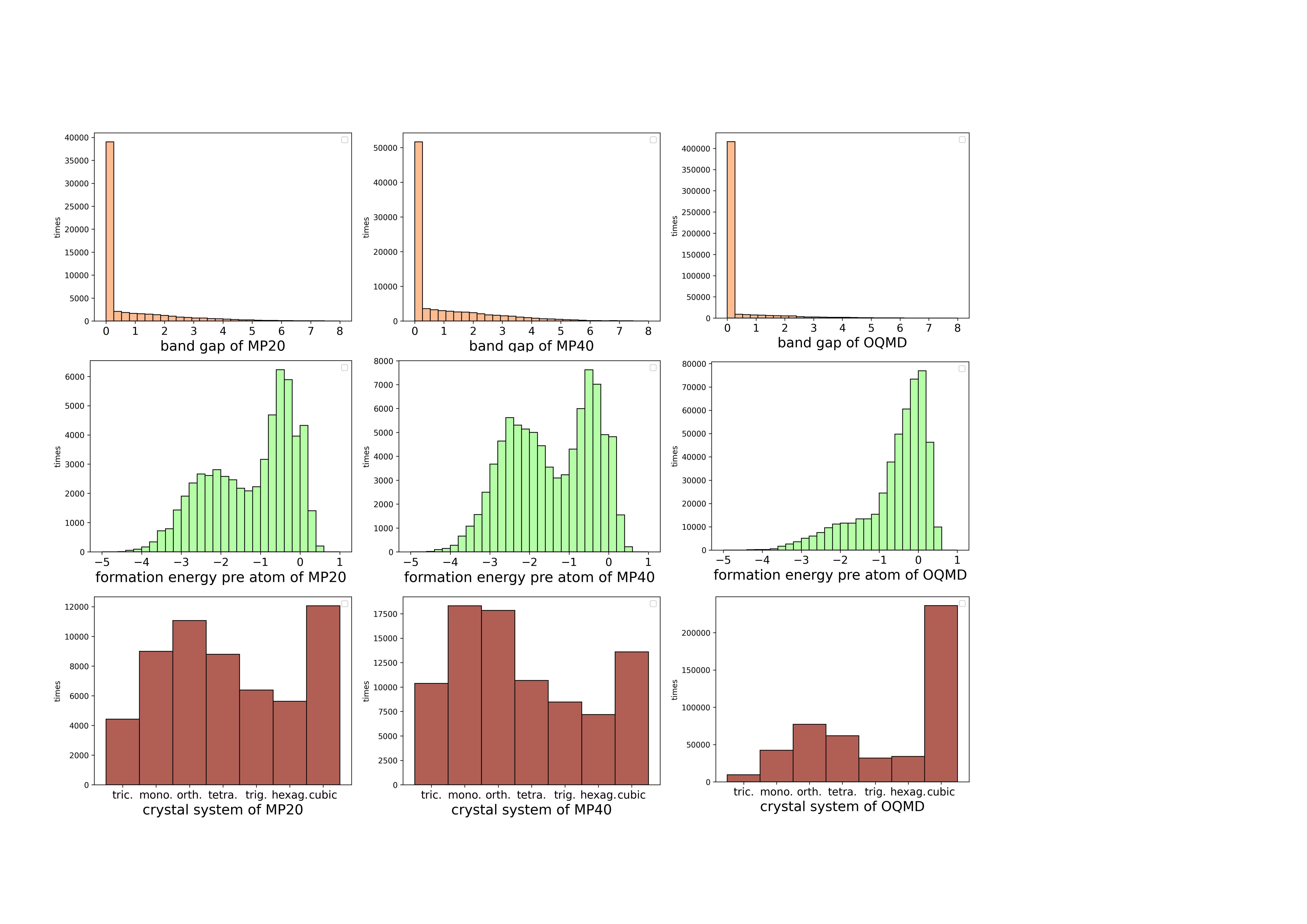}
	\caption{Distribution of band gap, formation energy and crystal system in three sub-datasets.}
	\label{fig:distribution}
\end{figure}

\autoref{fig:distribution} shows the distribution of the three data sets used in this paper in the three attributes of band gap, formation energy and crystal system. We can see that in the three data sets, crystals with zero-bandgap occupy a large part, and the data distribution in the formation energy is relatively wider. We think this is one of the important reasons why the conditional generation of the band gap looks worse than the conditional generation of the forming energy. 

In the comparison of OQMD and MP dataset, we found that the crystal formation energy in OQMD is mostly concentrated near 0eV, while the formation energy of MP crystal shows two peaks, which indicates that the average stability of crystal in MP is higher than that of OQMD, and also reflects that the vast material space is still not fully explored by human beings. In OQMD, the cubic crystal system is a large part of the material, while in MP it is relatively average.

\renewcommand\thetable{B\arabic{table}} 
\section*{Appendix B: Generate crystals with crystal system as condition}
\setcounter{table}{0}
\begin{table}[]
	\centering
	\renewcommand{\arraystretch}{1.0}
	\caption{The generation success rate of models trained with different data sets using different strategies is obtained by using different crystal systems as generation conditions}
	\begin{tabular}{cccccccc}
		\hline 
		target CS & tric. (\%)&mono. (\%)&orth. (\%)&tetra. (\%)&trig. (\%)&hexag. (\%)&cubic (\%) \\ 
		\hline 
		MP20-D  &   100.0 & 0.0 & 0.0 &2.0&0.0&0.0&0.0 \\ 
		MP20-F  &   100.0 & 29.0 &\textbf{ 83.0} &\textbf{74.0}&\textbf{81.0}&0.0&\textbf{93.0} \\ 
		MP20-L  &   100.0 & 0.0 & 0.0 &3.0&4.5&0.0& 31.0\\ 
		\hline 
		MP40-D  &   100.0 & 0.0 & 0.0 &0.0&3.0&0.0&0.0 \\ 
		MP40-F  &   100.0 & 0.0 & 3.0 &24.0&23.5&0.0&68.0 \\ 
		MP40-L  &   92.0 & 3.0 & 1.0 &0.0&1.0&0.0& 14.5\\  
		\hline 
		OQMD-D  &   86.0 & 0.0 & 22.0 &27.0&25.5&7.5&\textbf{93.5} \\ 
		OQMD-F  &   48.0 & \textbf{57.5} & 3.0 &\textbf{88.5}&\textbf{87.0}&0.0&\textbf{100.0} \\ 
		OQMD-L  &   100.0 & 0.5 & 37.5 &9.5&21.0&\textbf{22.0}& 2.0\\  
		\hline 
	\end{tabular}
	\label{tab:crystal-system}
\end{table}

\autoref{tab:crystal-system} shows the result of crystals generation with crystal system as condition. We see that the formation of triclinic systems seems good, but this is because many of the resulting crystals cannot be recognized as symmetrical. The important reason for this result is that the tolerance set by \texttt{pymatgen} when we identify crystal symmetry is 0.2 $\si{\angstrom}$ (twenty times the default value), and the RMSD before and after DFT relaxation is already higher than this tolerance, reaching 0.307 $\si{\angstrom}$. This suggests that the atomic positions of the crystals generated by our model need to be more precise. Another reason may be that the generation of crystal system and symmetry may require special design of the model in order to perform well under the current data volume and parameter number.

\section*{Appendix C: DFT parameter}
In this study, the relaxation of crystal structures is performed using the Vienna Ab initio Simulation Package (VASP-5.4.4)\cite{vasp}. Lattice shape, unit cell volume, and atomic positions are allowed to change. The PBE exchange-correlation functional\cite{PBE} is employed, and the KPOINTS file is generated using vaspkit\cite{vaspkit} with the Kmesh-Resolved Value set to 0.03 (2*PI/$\si{\angstrom}$). The cutoff energy for the plane wave basis set is set to 1.5 times the largest ENMAX value in the POTCAR file. The relaxation process is terminated when the norms of all atomic forces were smaller than 0.001 eV. A conjugate gradient algorithm was utilized to achieve the instantaneous ground state for the atoms.

\section*{Code availability}
Our code is avaliable at \href{https://github.com/cyye001/Con-CDVAE}{https://github.com/cyye001/Con-CDVAE}.

\section*{Note}
During development of the Con-CDVAE model, we become aware of a pre-print by Xie T et al that proposes a diffusion model\cite{MatterGen} with the same aim as our, but the method to realize conditional generation is different.

\addcontentsline{toc}{chapter}{Acknowledgment}
\section*{Acknowledgment}
This work was supported by the National Key Research and Development Program of China (Grant No. 2023YFA1607400, 2022YFA1403800), the National Natural Science Foundation of China (Grant No.12274436, 11925408, 11921004), the Science Center of the National Natural Science Foundation of China (Grant No. 12188101), and H.W. acknowledge support from the Informatization Plan of the Chinese Academy of Sciences (CASWX2021SF-0102) and the New Cornerstone Science Foundation through the XPLORER PRIZE.

\bibliography{main.bib}

\end{document}